\documentclass{aa}
\usepackage{listings}
\usepackage{graphicx}
\usepackage{txfonts}
\usepackage[multiple]{footmisc}

\newcommand\blfootnote[1]{%
  \begingroup
  \renewcommand\thefootnote{}\footnote{#1}%
  \addtocounter{footnote}{-1}%
  \endgroup
}

\begin{document}

   \title{The less and the more IGM transmitted galaxies from $z\sim2.7$ to $z\sim6$ from VANDELS\thanks{Based on observations made with ESO Telescopes at the La Silla or Paranal Observatories under programme ID(s) 194.A-2003} and VUDS\thanks{Based on data obtained with the European Southern Observatory Very Large Telescope, Paranal, Chile, under Large Program 185.A-0791}}
   \author{R. Thomas\inst{1}
           \and L. Pentericci\inst{\ref{RomeObs}} 
           \and O. Le F\`evre\inst{\ref{LAM},**}
           \and A. M. Koekemoer\inst{\ref{StSci}}
           \and M. Castellano \inst{\ref{RomeObs}}
           \and A. Cimatti\inst{\ref{Bologna}, \ref{Firenze}}
           \and F. Fontanot \inst{\ref{Trieste}, \ref{Triest2}}
           \and A. Gargiulo\inst{\ref{INAF-MILANO}}
           \and B. Garilli\inst{\ref{INAF-MILANO}}
           \and M. Talia \inst{\ref{Bologna},\ref{BolognaUniv}}
           \and R. Amorín\inst{\ref{serena1}, \ref{serena2}}
           \and S. Bardelli\inst{\ref{Bologna}}
           \and S. Cristiani\inst{\ref{Trieste}}
           \and G. Cresci, \inst{\ref{Firenze}}
           \and M. Franco\inst{\ref{CEA}, \ref{UK}}
           \and J. P. U. Fynbo\inst{\ref{dawn}, \ref{Niels} }
           \and N. P. Hathi\inst{\ref{StSci}}
           \and P. Hibon\inst{\ref{ESOChile}}
           \and Y. Khusanova\inst{\ref{LAM}, \ref{MPIE_hei}}
           \and V. Le Brun\inst{\ref{LAM}}
           \and B. C. Lemaux\inst{\ref{Davis}}
           \and F. Mannucci\inst{\ref{Firenze}}
           \and D. Schaerer\inst{\ref{geneve}}
           \and G. Zamorani\inst{\ref{Bologna}}
           \and E. Zucca\inst{\ref{Bologna}}
           }
\institute{European Southern Observatory, Av. Alonso de C\'ordova 3107, Vitacura, Santiago, Chile \label{ESOChile}, \\
              \email{rthomas@eso.org}
\and INAF - Osservatorio Astronomico di Roma, via Frascati 33, I-00078 Monteporzio Catone, Italy\label{RomeObs}
\and Aix Marseille Universit\'e, CNRS, LAM (Laboratoire d'Astrophysique de Marseille) UMR 7326, 13388, Marseille, France\label{LAM}
\and INAF - Osservatorio di Astrofisica e Scienza dello Spazio, via Gobetti 93/3, 40129 Bologna Italy\label{Bologna}
\and INAF - Osservatorio Astrofisico di Arcetri, largo E. Fermi 5, 50125, Firenze, Italy\label{Firenze}
\and INAF-Astronomical Observatory, via G.B. Tiepolo 11, I-34143 Trieste, Italy \label{Trieste}
\and IFPU - Institute for Fundamental Physics of the Universe, via Beirut 2, 34151, Trieste, Italy \label{Triest2}
\and INAF IASF-Milano, via A.Corti 12,20133, Milano, Italy\label{INAF-MILANO}
\and University of Bologna - Department of Physics and Astronomy, Via Gobetti 93/2, I-40129, Bologna, Italy\label{BolognaUniv}
\and Instituto de Investigaci\'on Multidisciplinar en Ciencia y Tecnolog\'ia, Universidad de La Serena, Ra\'ul Bitr\'an 1305, La Serena, Chile\label{serena1}
\and Departamento de Astronom\'ia, Universidad de La Serena, Av. Juan Cisternas 1200 Norte, La Serena, Chile\label{serena2}
\and AIM, CEA, CNRS, Universit\'{e} Paris-Saclay, Universit\'{e} Paris Diderot, Sorbonne Paris Cit\'{e}, F-91191 Gif-sur-Yvette, France\label{CEA}
\and Centre for Astrophysics Research, University of Hertfordshire, Hatfield, AL10 9AB, UK \label{UK}
\and Cosmic DAWN Center, Vibenshuset, Lyngbyvej 2, DK-2100 Copenhagen, Denmark \label{dawn}
\and Niels Bohr Institute, Copenhagen University, Lyngbyvej 2, DK-2100 Copenhagen O, Denmark \label{Niels}
\and Space Telescope Science Institute, 3700 San Martin Drive, Baltimore, MD 21218, USA\label{StSci}
\and Max-Planck-Institut f\"{u}r Astronomie, K\"{o}nigstuhl 17, D-69117 Heidelberg, Germany\label{MPIE_hei}
\and Department of Physics, University of California, Davis, One Shields Ave., Davis, CA 95616, USA \label{Davis}
\and Observatoire de Gen\`eve, Universit\'e de Gen\`eve, 51 Ch. des Maillettes, 1290 Versoix, Switzerland \label{geneve}
          }

  \date{Received ; accepted }

 
  \abstract
   {}
   {Our aim is to analyse the variance of the Inter-Galactic Medium transmission (IGM) by studying this parameter in the rest-frame UV spectra of a large sample of high redshift galaxies.}
   {We make use of the VIMOS Ultra Deep Survey and the VANDELS public survey to have an insight into the far UV spectrum of $2.7<z<6$ galaxies. Using the SPARTAN fitting software, we estimate the IGM towards individual galaxies and then divide them in two sub-samples characterized by a transmission above or below the theoretical prescription. We create average spectra of combined VUDS and VANDELS data for each set of galaxies in seven redshift bins.}
   {The resulting spectra clearly exhibit the variance of the IGM transmission that can be seen directly from high redshift galaxy observations. Computing the optical depth based on the IGM transmission, we find an excellent agreement with QSOs results. In addition, our measurements seem to suggest that there is a large dispersion of redshift where complete Gunn-Peterson Trough happens, depending on the line of sight.}
   {}


\authorrunning{R. Thomas et al }
\titlerunning{The less and the more IGM transmitted galaxies from $z\sim2.7$ to $z\sim6$}

   \maketitle
%

\section{Introduction}
\label{intro}

\blfootnote{$^{**}$ Oliver Le F\`evre, 1960-2020}

The light coming from distant sources (like Quasi Stellar Objects and galaxies) is absorbed by the gaseous hydrogen systems that are lying along the line of sight. At increasing redshift, this effect, called the Inter-Galactic Medium (IGM) absorption can be so important that all the light at wavelengths blueward the Lyman $\alpha$ [Ly$\alpha$] at 1216\AA~ becomes invisible to us. Such absorption has been the topic of numerous studies and this is mainly connected with the growth of the Large Scale Structure as predicted by the hierarchical structure formation scenario and it is tightly connect with the epoch of Reionization (e.g. \citealt{cen94}). 

Twenty five years ago, Madau (\citealt{madau95}, hereafter \textit{M95}) proposed a theoretical model able to reproduce the shape of the extinction curve as a function of redshift. Using this model, they conclude that the IGM transmission, Tr(Ly$\alpha$), decreases with increasing redshift and that the scatter, at a given redshift should be large, e.g., from 20\% to 70\% with an average of 40\% at z=3.5. A decade later, Meiksin (\citealt{Meiksin06}, hereafter \textit{M06}) produced an update of this model using the $\Lambda$ Cold Dark Matter cosmological predictions. More recently, \cite{Inoue14} developed a new model of transmission. Their model predicts a weaker average absorption in the range z=3-5 while it becomes stronger at z>6. On the observational side, studies of the IGM have been provided almost exclusively by Quasi Stellar Objects (QSOs). Faucher-Giguère et al. (2008b) used 86 high-resolution quasar spectra with a high signal-to-noise ratio to provide reference measurements of the dispersion over 2.2 < z < 4.6.\citealt{dallaglio08} used 40 quasars bright quasars to produce the measurements of the IGM optical depth between z=2.5 and z=4.5 while \citet{Becker13} has extended the redshift range up to 5.5 with more than 6000 quasars. They all found that the transmission is indeed decreasing with increasing redshift. 

Until a few years ago, no observational studies had been made of the evolution of the IGM transmission from galaxy samples mainly because of the lack of large spectroscopic samples with high signal-to-noise ratios at high redshift that probed a wavelength range between the Lyman-limit at 912\AA~ and Ly$\alpha$. Hence, the IGM transmission towards extended galaxies had not yet been compared with the well known results for point-like QSO samples. In two recent papers we were able to compute the IGM transmission for a large sample of galaxies using the VIMOS Ultra Deep Survey [VUDS; \citealt{OLF15, Thomas17}, hereafter T17] and the VANDELS survey [\citealt{VANDELS, Pent18, Thomas20}, hereafter T20]. In these studies we showed that not only we can measure the IGM transmission using galaxy spectra but that it is crucial for understanding how galaxies are selected with Lyman Break techniques \citep{Steidel95} which highly depends on the IGM.

In this letter, we assembled more than 2500 galaxies from VUDS and VANDELS at z>2.7 with measured IGM absorption (using the SPARTAN tool) to create averaged IGM spectra in different redshift bins to look at the IGM extinctions and visually see the effect of the variance of IGM on galaxy data. We describe the VUDS and VANDELS galaxy sample and selection in Sect~\ref{Data} . The fitting method with the SPARTAN tool \citep{SPARTAN} and the recipe we use to compute the IGM are described in Sect~\ref{SPARTAN}. The results are presented in Sect~\ref{results} and Sect.~\ref{optdept}. We discuss the results in Sect~\ref{disc} and conclude in Sect.~\ref{conc}.

All magnitudes are given in the AB system \citep{Oke83} and we use a cosmology with $\Omega_M = 0.3$, $\Omega_{\Lambda} = 0.7$ and $h = 0.7$.

\section{Data}
\label{Data}

\begin{table*}[h!]
\centering
\caption{Details of the stacked spectra displayed in Fig. \ref{stacks}. For each stack we provide the number of individual spectra entering the stacking of the spectrum (N$_X$) and the average redshift of all these galaxies ($<z_X>$). 
Quantities with `all', `more', `less' refer to the stack with all the galaxies (in grey in Fig.\ref{stacks}), galaxies with the more transmitting IGM (blue) and galaxies with the less transmitting IGM (red). The last but one column shows the flux difference between the more IGM transmitted galaxies and the less IGM transmitted galaxies, computed at 1070-1170\AA~while the last column shows the fractional difference.  It is worth mentioning that $N_{all} \neq N_{more}$+$N_{less}$ as we are considering object $0.5\sigma$ above ($N_{more}$) or below ($N_{less}$) the average. $N_{all}$ include as well all galaxies with an IGM absorption equal to the theoretical average. Finally, the Signal to noise ratio (SNR), is the one computed in the stacked spectra in the IGM region at 1070-1170\AA.}
\begin{tabular}{ccccccccccccc}
\hline
Redshift bin & <z$_{all}$> & $N_{all}$ & <z$_{more}$>  & $N_{more}$ & SNR$_{more}$ & <z$_{less}$> & $N_{less}$ & SNR$_{less}$ & $\Delta <$F$>_{1070-1170}$ & $\frac{\Delta <F>}{<F_{all}>}$ \\ 
\hline
$2.7<z<3.0$ & 2.8583 & 749 & 2.8643 & 333 & 17.8 &2.8548 & 197 & 11.2& 0.37 & 0.45\\
$3.0<z<3.3$ & 3.1392 & 546 & 3.1225 & 188 & 17.3 &3.1540 & 183 & 13.4&0.31 & 0.41\\
$3.3<z<3.6$ & 3.4315 & 412 & 3.4285 & 102 & 15.8 &3.4361 & 181 & 11.6&0.29 & 0.44\\
$3.6<z<4.0$ & 3.8133 & 305 & 3.7901 & 80  & 13.6 &3.8146 & 105 & 8.2 &0.23 & 0.36\\
$4.0<z<4.5$ & 4.2198 & 274 & 4.2545 & 113 & 14.2 &4.1693 & 53  & 7 &0.28 & 0.44\\
$4.5<z<4.8$ & 4.6209 & 122 & 4.6157 & 58 & 8.8 &4.6472 & 16 & 2.7 &0.24 & 0.42\\
$z>4.8$     & 5.0411 & 94 & 5.0707 & 45  & 5 &5.0464 & 18 & 2.6&0.14 & 0.34\\
\hline 
\end{tabular} 
\label{Table_avail}
\end{table*}

In this paper, we use deep spectra collected by the VUDS \citep{OLF15} and VANDELS \citep{VANDELS, Pent18}. We briefly describe here both surveys and refer the reader to the survey description papers for more in-depth details. Both surveys have been carried out with the now-decommissioned VIsible Multi-Object Spectrograph [VIMOS] instrument installed at the Nasmyth focus of Unit Telescope 3 of the Very Large Telescope.

VUDS (P.I. O. Le F\`evre) is a European Southern Observatory [ESO] large program designed to study the first billion years of galaxy evolution. It is based on the observation of ten thousand sources up to $z=6.5$ selected primarily via the photometric redshift method performed with the LePhare software \citep{Arnouts99, Ilbert06}. Each target was observed for a total of $\sim14$h in the wavelength range $3500\leq\lambda\leq9350$\AA~ using the two low resolution grism of the instrument at R$\sim$240. The data reduction was carried out using the VIPGI software \citep{Sco05}. As the targets are in well known fields (VIMOS VLT Deep Survey [VVDS] 2h field, Cosmic evolution Survey [COSMOS] and Extended Chandra Deep Field South [ECDFS]), a large amount of multi-wavelength photometric data are available to complement the spectroscopic information. We direct the reader to \citet{OLF15} for a full description of the survey.

The second sample comes from the VANDELS survey (P.Is R. McLure \& L. Pentericci). VANDELS is an ESO public spectroscopic survey aiming at the exploration of the high redshift Universe at $1<z<7$. More than 2100 galaxies have been observed using the VIMOS medium resolution grism at R$\sim$600 from 4800 to 10000 \AA~and the exposure time varied from 20h up to 80h. As it was the case for VUDS, most targets were selected using the photometric redshift technique. The data reduction was carried out using the EASYLIFE package \citep{EASYLIFE}. Targets were selected in the two widely observed Ultra deep survey [UDS] and Chandra deep field south [CDFS] fields which allows the association of the spectroscopy to a large amount of multi-wavelength photometric data. Readers can find more information about the survey in the two papers presenting it: \citet{VANDELS} and \citet{Pent18}.\\
For both surveys, all the redshift estimations were performed using the EZ software \citep{Gari10}, a template cross-correlating code, and they both use an equivalent redshift flag system to assess the quality of the redshift measurements. In this scheme, galaxies with a redshift-flag 2,3,4 are those with the most reliable redshift measurements, with a probability to be correct of 75\%, 95\%, and 100\%, respectively.

In this study, we are interested in galaxies with IGM transmissions higher and lower than the mean theoretical transmission. As defined in T17 and T20, this quantity is measured between rest-frame 1070\AA~and 1170\AA. Such requirement imposes a lower redshift limit of z=3.7 for VANDELS (taking a lower wavelength of $\sim$5000\AA) and z=2.7 for VUDS (lower wavelength at $\sim$3900\AA). Finally, we select only galaxies with a redshift flag of 2, 3 or 4 which ensures to have the best quality redshift. This leads to a combined sample of 2502 galaxies including 296 galaxies from VANDELS and 2206 galaxies from VUDS. As the goal is to analyse galaxy spectra, it is worth mentioning that VANDELS spectra have been rebinned to match VUDS wavelength grid.


\begin{figure*}[h!]
\centering
\includegraphics[width=6.cm]{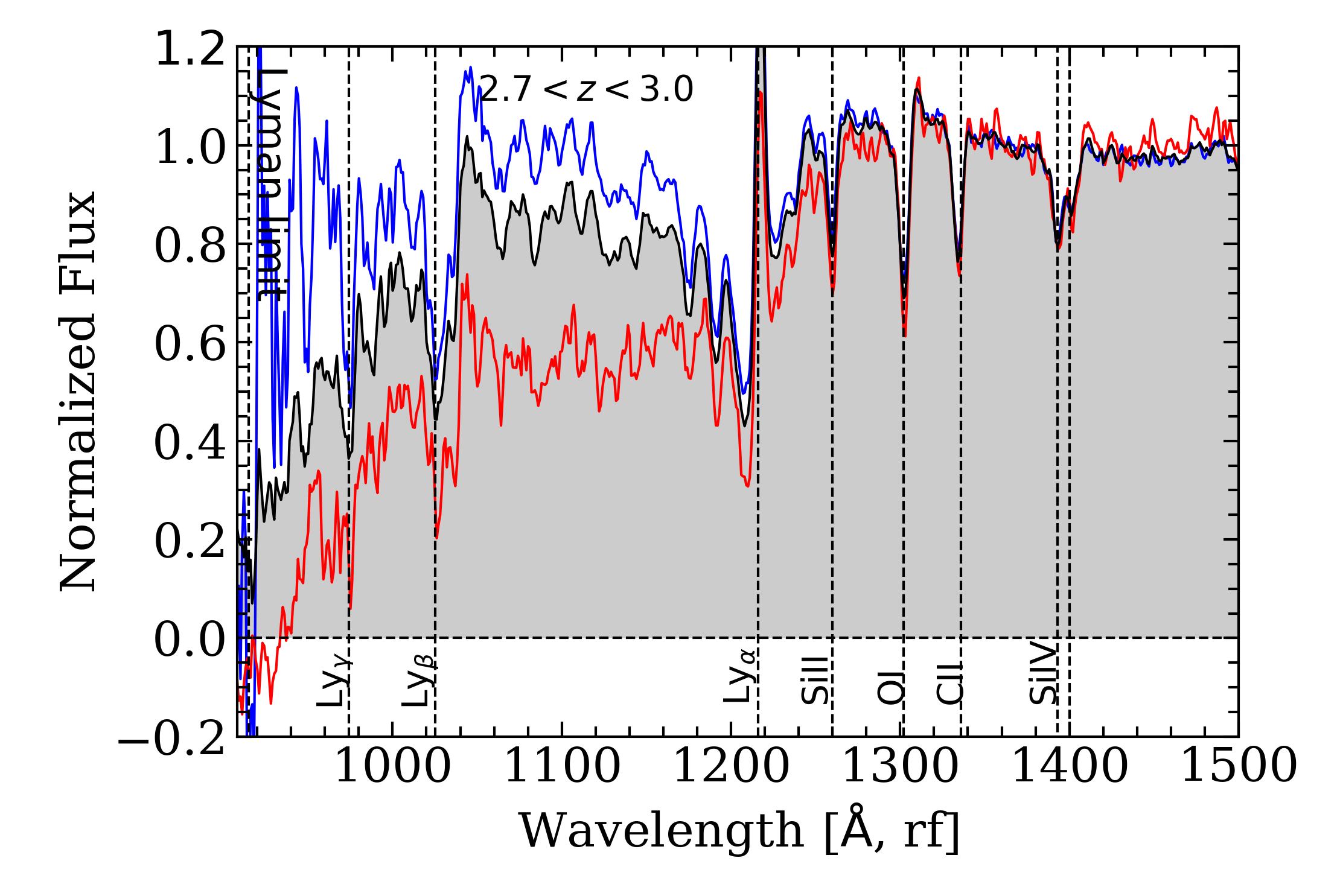}
\includegraphics[width=6.cm]{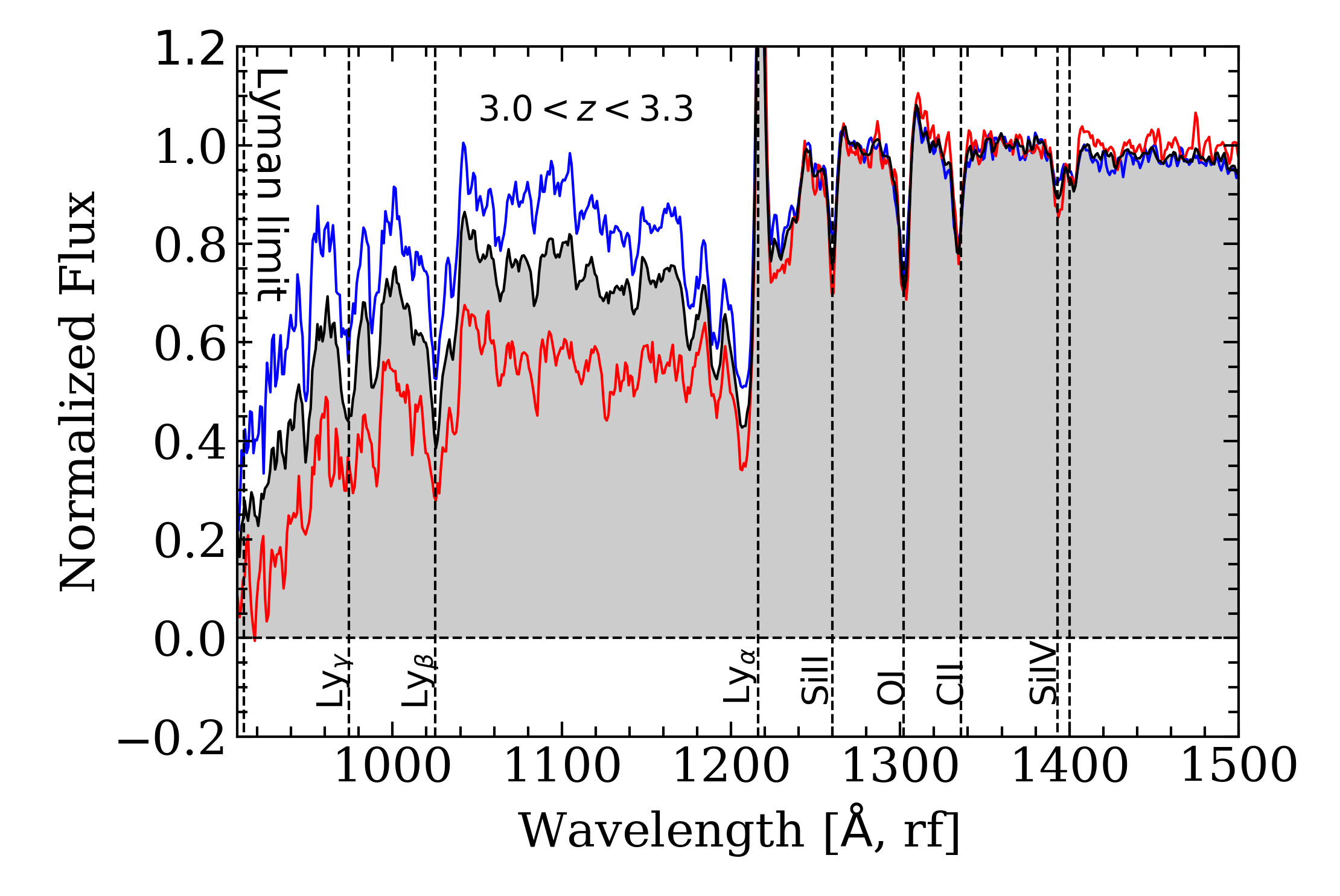}
\includegraphics[width=6.cm]{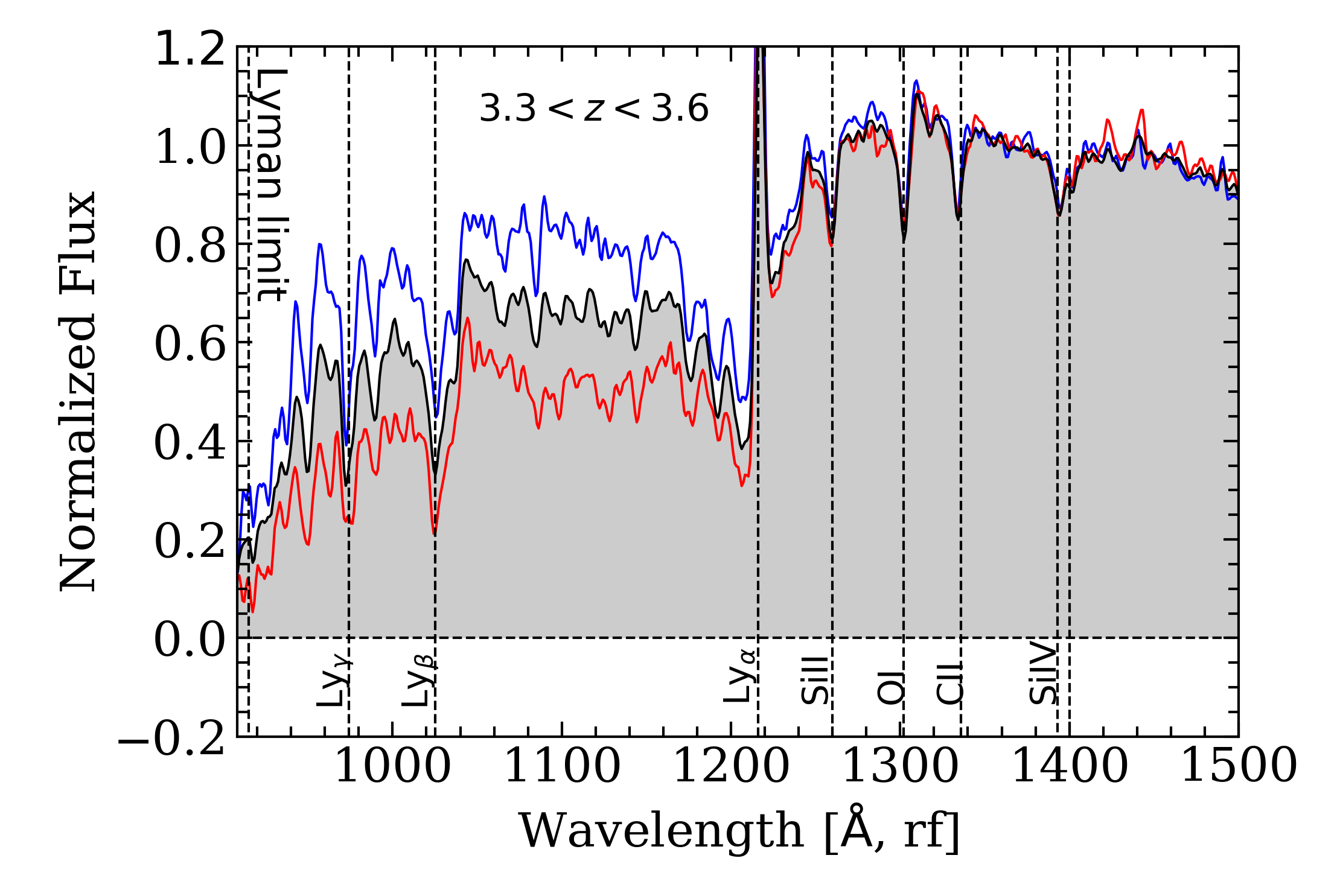}
\includegraphics[width=6.cm]{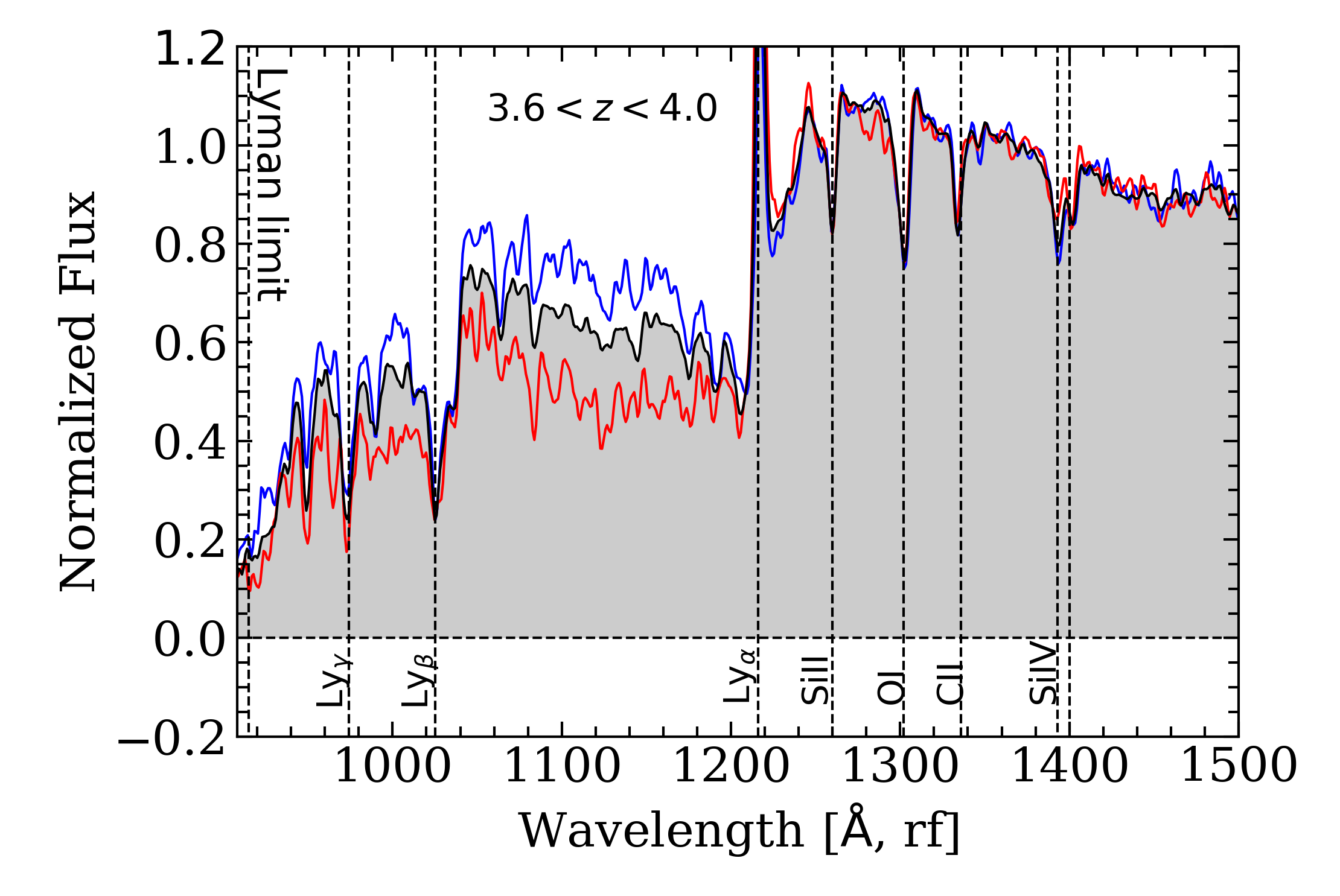}
\includegraphics[width=6.cm]{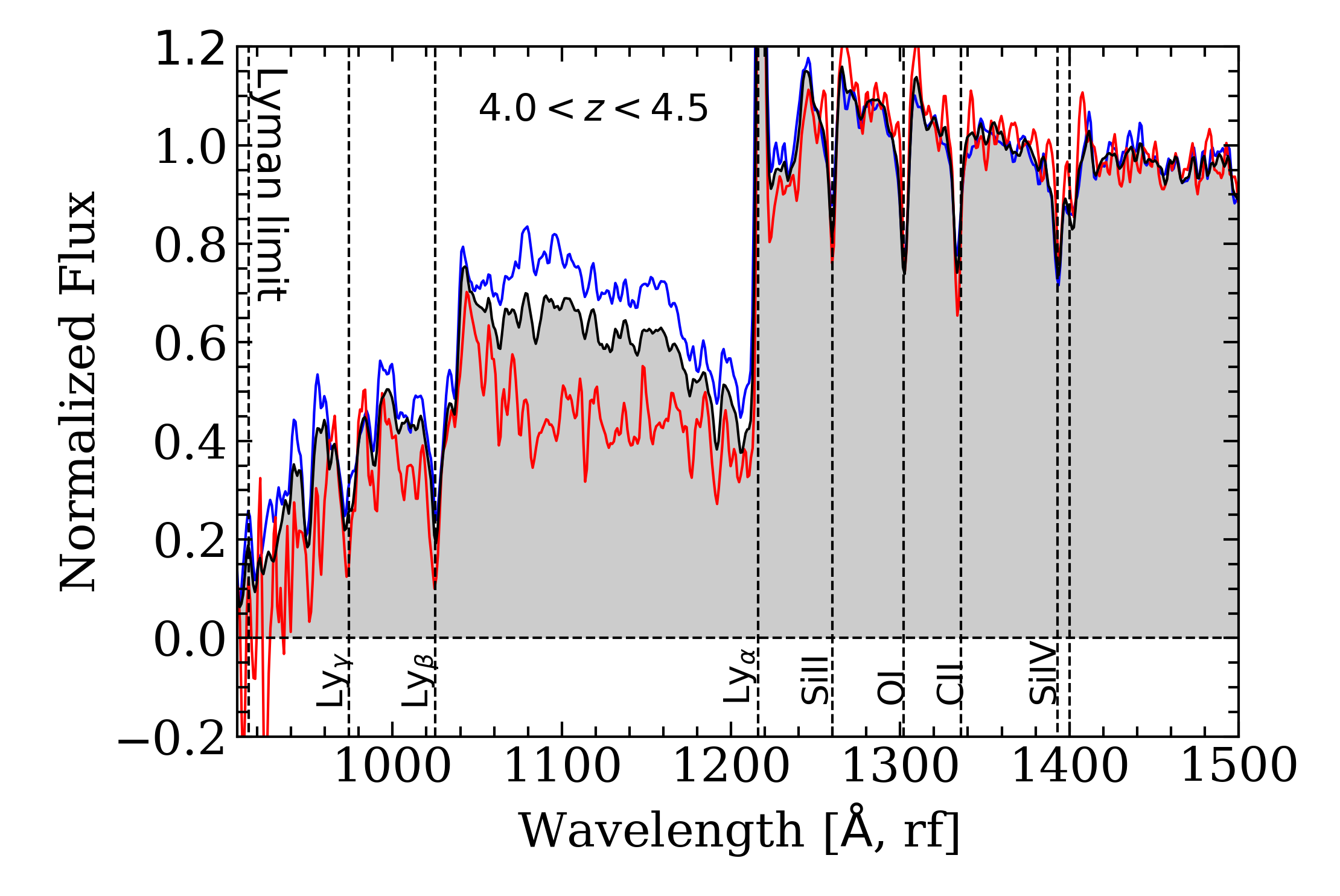}
\includegraphics[width=6.cm]{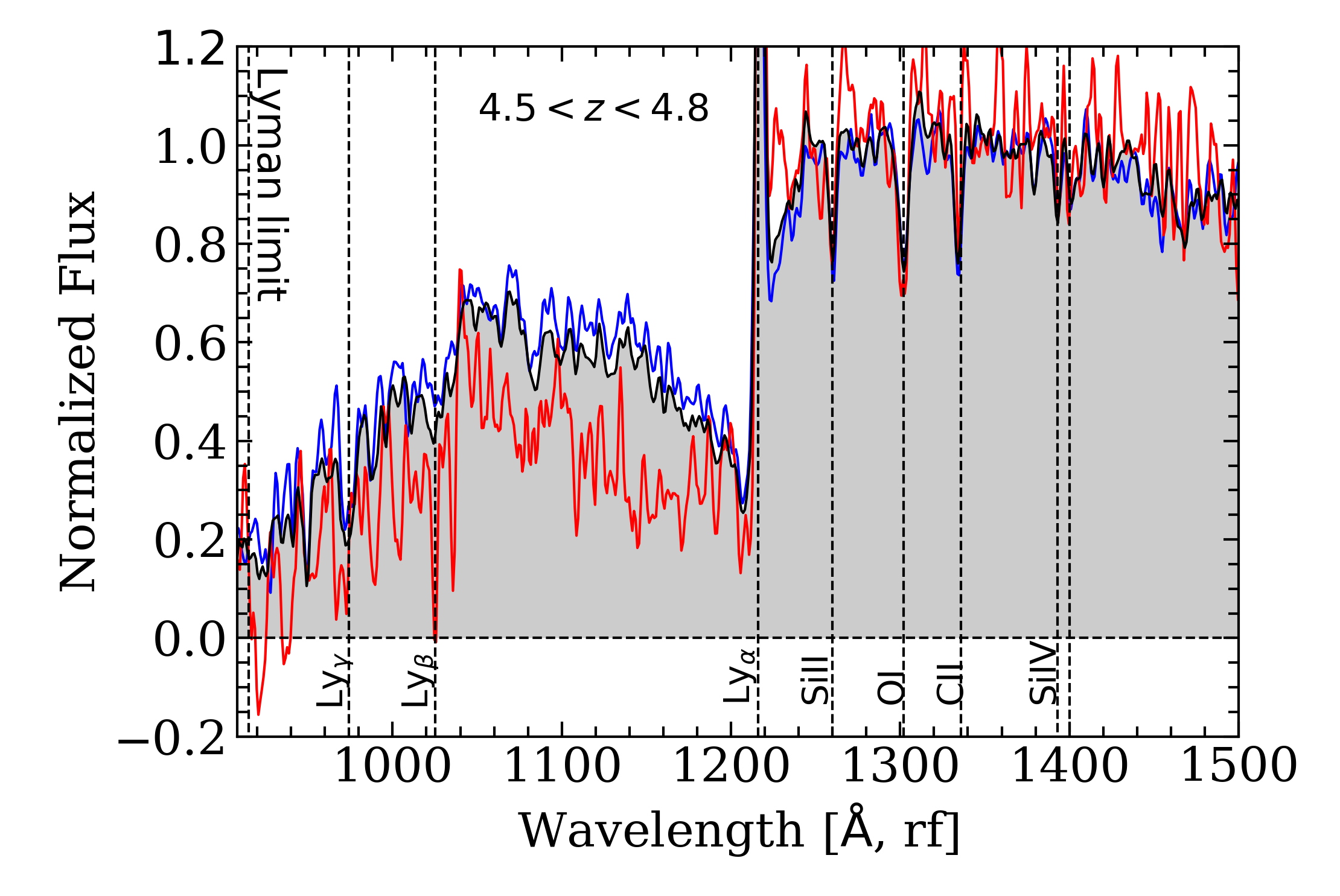}
\includegraphics[width=6.cm]{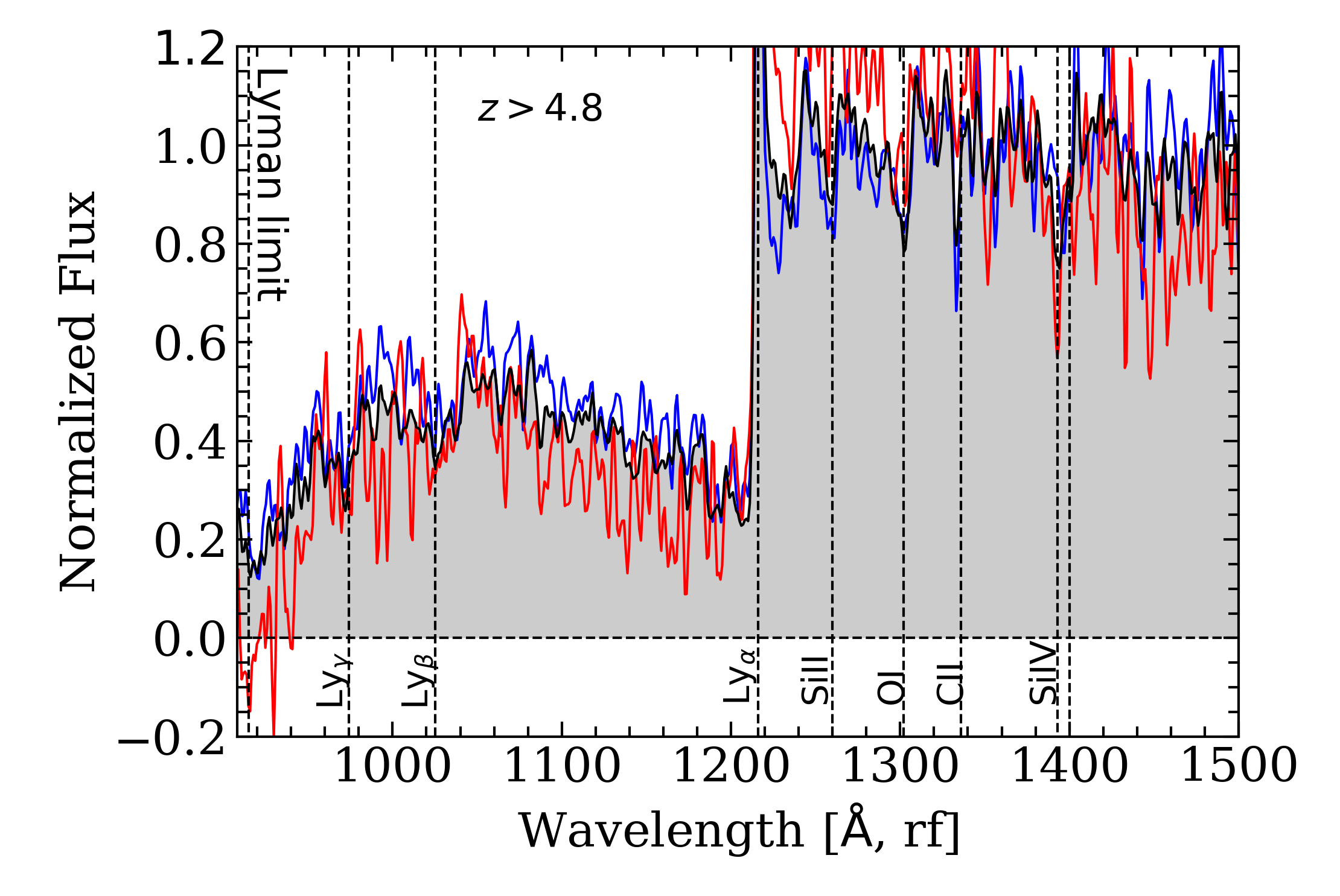}
\caption{Average spectra with different IGM-attenuated sample. All spectra are given in the restframe (rf). We show 7 different redshift bins (from top left to bottom left): $2.7<z<3.0$, $3.0<z<3.3$, $3.3<z<3.6$, $3.6<z<4.0$, $4.0<z<4.5$, $4.5<z<4.8$ and $z>4.8$. For each panel we show the averaged spectrum for the more transmitted galaxies in blue, the averaged spectrum of the less transmitted galaxies in red and the averaged spectrum of all the galaxies in grey}
\label{stacks}
\end{figure*}

\section{Method: The SPARTAN tool and IGM estimation}
\label{SPARTAN}

The estimation of the IGM transmission is done using the SPARTAN (Spectroscopic And photometRic fitting Tool for Astronomical aNalysis) software \citep{SPARTAN}. It is a Python software created to be able to fit both photometry and spectroscopy in a common environment. We provide in this section the relevant aspects for this paper. SPARTAN uses a $\chi^{2}$ minimization technique that consist on a comparison of the observations to theoretical galaxy models (see T20 for more details).


In order to select galaxies on the basis of their estimated IGM transmission, we need to measure it. This is done through the two-step template fitting method described in T20 which allows for the reduction of the degeneracy between dust extinction and IGM absorption. The first step is done to estimate the dust attenuation parameter E(B-V)$_{s}$ based on the SPARTAN fitting of all the photometric data point only. The second step is done fixing the dust extinction (from the first pass)  on the spectral data which provides the IGM transmission. The fitting is done using \cite{BC03} models with a \cite{Chab03} initial mass function. The stellar-phase metalicity ranges from sub-solar (0.2$Z_{\odot}$ and 0.4$Z_{\odot}$) to solar (1.0$Z_{\odot}$). We assume an exponentially delayed star formation history of the form $SFR \propto t \times \tau^{-2} \times \exp(-t/\tau)$ with a time-scale parameter, $\tau$, ranging from 0.1 Gyr to 2.0 Gyr. The ages (corresponding to the time since onset of star formation) range from 0.01 Gyr to 4 Gyr. It is worth noting that this range of age is further limited by the age of the Universe at the redshift that is considered during the fit. For the photometric fitting, the E(B-V)$_{s}$ parameter can vary from 0.0 to 0.39 (in 0.03 steps). As described above, this parameter is fixed during the second pass spectral fitting. Finally, the IGM prescription that is used from T17 in which we built IGM transmission templates around the mean of M06. At a given redshift, six empirical additional curves at $\pm0.5\sigma$, $\pm1.0\sigma$ and $\pm1.5\sigma$ were created. This allows us to use the IGM as a free parameter in our fit and explore a large range of IGM transmission.  At $z=3$ the IGM transmission, Tr(Ly$\alpha$), can range from 20\% to 100\% while at $z=5$, it might vary from 5\% to 50\%. The Ly$\alpha$ transmission is estimated directly on these models as the average of the selected model between 1070\AA~and 1170\AA~(we refer the reader to T20 for examples of spectral fitting).

\section{The more and less IGM transmitted galaxies at z>2.7}

\label{results}
\begin{figure*}[h!]
\centering
\includegraphics[width=17.5cm]{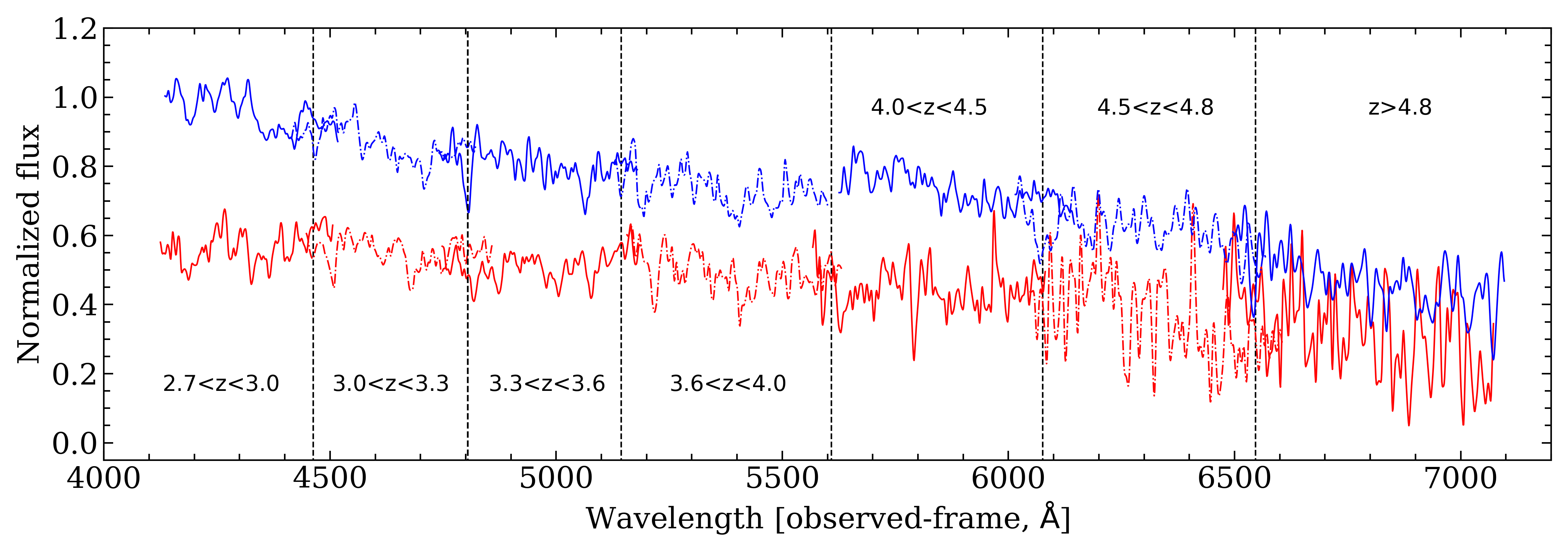}
\caption{Evolution of the region where the Ly$\alpha$ transmission is computed (1070-1170\AA) in the different redshift bins of Fig.\ref{stacks} but this time in the observed frame (the redshift of the spectra are set to the average redshift of the redshift bin). The vertical dashed lines show the separation between each redshift bin (in the middle of the overlap). The color coding is the same as Fig.\ref{stacks}. }
\label{stacks_observed}
\end{figure*}

The objective of this paper is to look at the galaxies in different redshift bins which have an IGM transmission different from the theoretical mean. In order to achieve this goal, we create average spectra of galaxies selected based on their IGM transmission computed from the SPARTAN fitting. This has been done in seven redshift bins, $2.7\leq z<3.0$, $3.0\leq z<3.3$, $3.3\leq z<3.6$, $3.6\leq z<4.0$, $4.0\leq z<4.5$, $4.5\leq z<4.8$ and $z>4.8$. The choice of these seven bins was a trade-off between maximizing the number of points and keeping a high number of galaxies to achieve a SNR high enough for our study. Nevertheless, It is worth mentioning that the change of this binning, e.g. with lass bins, does not change our results. The galaxies in each bin were separated into two sub-samples: the more transmitted galaxies, i.e. those with an IGM transmission measured at a level higher than the predicted mean of the M06 prescription (with a selected model at $\geq0.5\sigma$, see previous section) and the less transmitted galaxies, i.e., with an IGM transmission measured at a level lower than the mean of the M06 prescription (with a selected model at $\leq0.5\sigma$). For each set of galaxies we have created an average spectrum using the \textit{specstack}\footnote{https://specstack.readthedocs.io/en/latest/} tool \citep{specstack}. The tool works as follows. For a given set of galaxies, we de-redshift all the individual spectra and normalise them in a region redward of the Ly$\alpha$ line (in our case 1345\AA-1380\AA~due to the absence of spectral feature in that region). Then we regrid the spectra in a common wavelength grid. Finally, at a given wavelength, we compute the average of all the flux density in each given pixel weighted by the individual spectra SNR using a 3$\sigma$ clipping method. In the seven redshift bins, the resulting average spectra based on our 2502 galaxies from both VUDS and VANDELS are presented in Fig.~\ref{stacks} and their associated data are displayed in Table \ref{Table_avail}. The averaged spectra were computed between 912\AA~and 1500\AA~rest-frame. In each redshift bin we show three average spectra: an average spectrum made from the more IGM-transmitted galaxies, an average spectrum made from the less IGM-transmitted galaxies and finally an average spectrum made from all the galaxies available in the considered redshift bin independent of the IGM transmission\footnote{The evolution of the Ly$\alpha$ transmission itself is shown in T20.}.

First of all, it is important to note that 70\% of the average spectra are made of more than 100 individual spectra, the other 30\% are at the highest redshift bins where we have less spectra. This results in the average spectrum corresponding to the less IGM transmitted galaxies having a higher level of noise. In each redshift bin, the average reshift of that comprise the three composite spectra are similar and it is therefore meaningful to compare them directly. Very importantly, Fig.\ref{stacks} shows that redward, of the Ly$\alpha$ line, the spectra of the three samples are very close to each other which means that the underlying galaxy populations are similar, more notably, the dust properties. Therefore, the strong variation of flux below the Ly$\alpha$ line depends primarily on the IGM extinction along the line of sight. At every redshift, the less transmitted sample is much dimmer at 1070\AA-1170\AA~than the more transmitted one, except for the highest redshift spectra (see next section). Based on the measurement on the average spectra, the flux density at 1070\AA-1170\AA~of less transmitted galaxies is on average 40\% fainter than that of more transmitted galaxies. This difference will have a very important influence on the selection of galaxies based on the Lyman-break technique (as shown in T17 at z$\sim$3.2 with \textit{ugr} colour-colour diagram). However, we note that for the two highest redshift bins, the flux density of the average sample (in gray) is very similar to that of more transmitted sample. In these bins, there are fewer galaxies in the sample which is less transmitting, therefore the average tends to be closer to the more transmitted one. We investigate the potential origin of this asymmetry in section \ref{sec_asym}. For a better visualisation, the evolution of the flux itself in this region is displayed in Fig.\ref{stacks_observed} where all the regions have been redshited to the average redshift of the bin (see Table~\ref{Table_avail}). 

\begin{figure}[h!]
\centering
\includegraphics[width=8.cm]{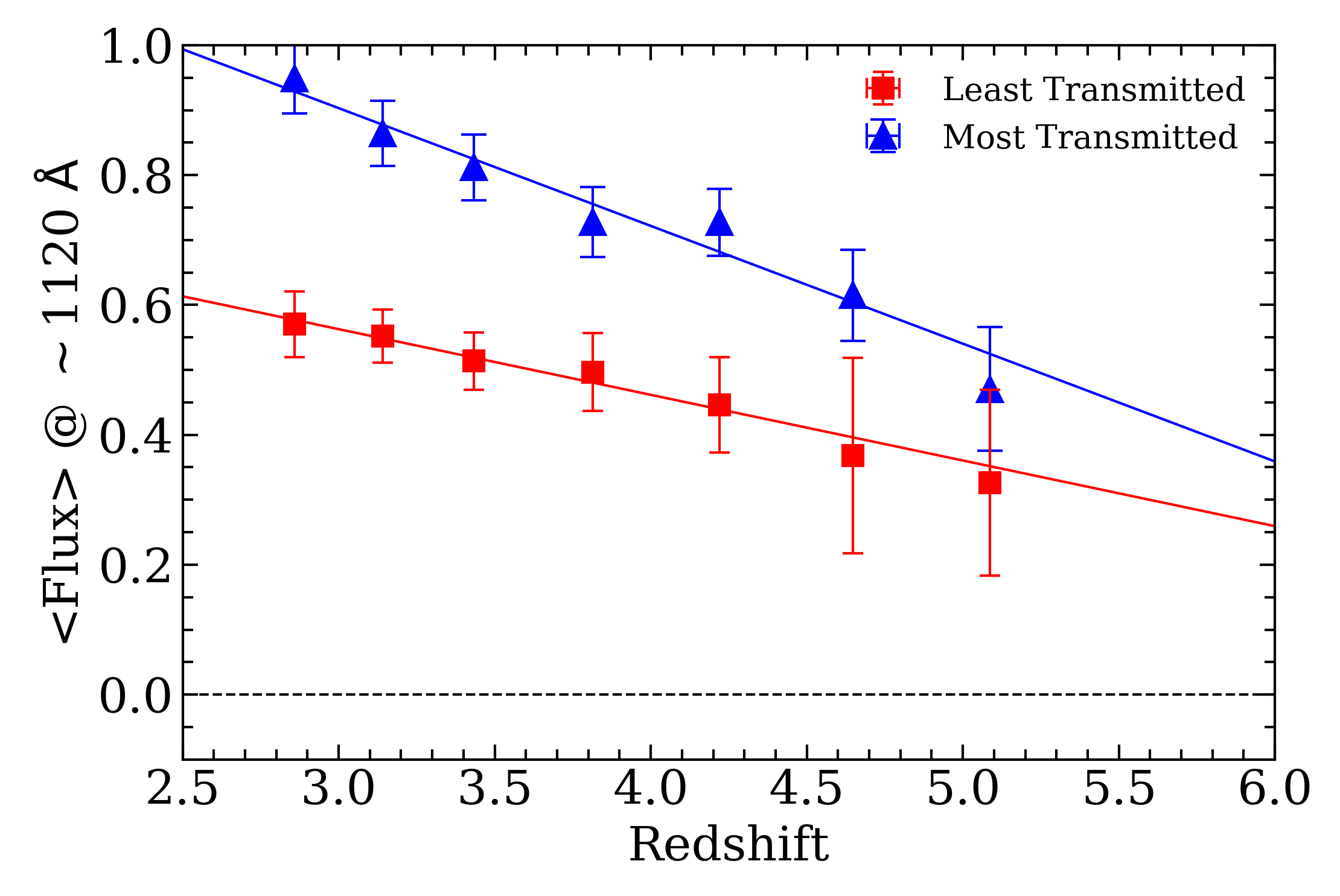}
\caption{Evolution of the averaged flux in the 1070-1170\AA~region in the averaged spectra for the more (in blue) and the less (red) IGM-transmitted average spectra. The two solid lines are linear fits to the evolutions whose functional shapes are given in equation~\ref{linearfit}.}
\label{ev}
\end{figure}

We compute the average flux of the rest-frame stacked spectra in the IGM transmission region (1070-1170\AA), and the evolution of this average flux is shown in Fig.\ref{ev}. As seen in the stacks, the average flux is decreasing with redshift as expected due to the evolution of the IGM (M06 \& M95). Nevertheless, the evolution seems to be different for the two samples. We compute the difference in average flux between the more and the less transmitted sample, $\Delta F$. These measurements are reported in the last but one column of Table \ref{Table_avail}. We see that the difference is decreasing with increasing redshift. It goes from 0.37 at z$\sim$2.85 down to 0.14 at z$\sim$5.04, indicating that the difference in transmission between the two sets of galaxies is reducing as well. We also computed the fractional difference, $\Delta F/<F_{all}>$ and reported it in the last column of Table \ref{Table_avail}. It slightly decreases with redshift going from 0.45 at z$\sim$2.85 to 0.34 at z$\sim$5.04. Nevertheless, the scatter of the points does not let us draw any conclusion from this evolution. These evolutions are well fitted by a simple linear function, $f(z)= a\times z +b$ (valid for taking all the points of Fig.\ref{ev} from z$_{source}$=2.85 and z$_{source}$=5.04):

\begin{equation}
\label{linearfit}
   \begin{split}
 <F>_{less} & = -0.101[\pm0.008] \times z + 0.866 [\pm0.029]  \\ 
 <F>_{more} & = -0.181[\pm0.019]  \times z + 1.447 [\pm0.072]    
   \end{split}
\end{equation}

The fitting method has been carried out using the curve\_fit function of the scipy.optimize package \citep{2020SciPy-NMeth} which performs a $\chi^{2}$ minimization. The errors on the parameters correspond to the standard deviation estimated from the covariance matrix\footnote{https://docs.scipy.org/doc/scipy/reference/generated/\\scipy.optimize.curve\_fit.html}.

\section{Optical depth}
\label{optdept}

QSO studies traditionally compute the HI optical depth from the IGM transmission defined as:
\begin{equation}
    \tau_{eff} = -\ln \mathrm{Tr(Ly_{\alpha})},
\end{equation}
where the Tr(Ly$_{\alpha}$) is the Lyman $\alpha$ transmission computed directly from the IGM template.
We apply this transformation to each of our set of galaxies and the results are reported in Table.~\ref{Table_optdep}. It is also worth mentioning that QSO studies often use the redshift of the absorbers in the Lyman $\alpha$ forest defined as:

\begin{equation}
\label{redtrans}
    \lambda_{Ly\alpha}(1+z_{abs}) = \lambda_{0}(1+z_{s}),
\end{equation}
with $z_s$ the redshift of the source and $\lambda_{Ly\alpha}=1215.67$\AA. To be able to compare our measurements to the QSO's literature, we transform all our redshifts in this section to $z_{abs}$ using $\lambda_{0}=1120$\AA, the middle of the region where we measure Tr(Ly$_{a}$).

\begin{table}[h!]
\centering
\caption{Optical depth from the IGM transmission in each of our galaxy subsamples.Quantities with `all', `more', `less' refer to the stack with all the galaxies (in grey in Fig.\ref{stacks}), galaxies with the more transmitting IGM (blue) and galaxies with the less transmitting IGM (red). Redshift have been transformed using eq.\ref{redtrans}.}
\begin{tabular}{cccccc}
\hline
$z_{abs}^{all}$ & $\tau_{eff}^{all}$  & $z_{abs}^{more}$ & $\tau_{eff}^{more}$  & $z_{abs}^{less}$ & $\tau_{eff}^{less}$\\ 
\hline
2.55 & $0.21\pm0.1$ & 2.56 & $0.01\pm0.05$ & 2.55 & $0.54\pm0.04$\\
2.81 & $0.30\pm0.1$ & 2.79 & $0.06\pm0.02$ & 2.83 & $0.61\pm0.04$\\
3.08 & $0.41\pm0.12$ & 3.07 & $0.13\pm0.03$ & 3.09 & $0.70\pm0.05$\\
3.43 & $0.56\pm0.15$ & 3.41 & $0.23\pm0.09$ & 3.44 & $0.86\pm0.07$\\
3.80 & $0.64\pm0.15$ & 3.84 & $0.41\pm0.09$ & 3.76 & $0.96\pm0.05$\\
4.17 & $0.87\pm0.19$ & 4.17 & $0.64\pm0.07$ & 4.20 & $1.23\pm0.06$\\
4.56 & $1.16\pm0.22$ & 4.59 & $0.91\pm0.12$ & 4.57 & $1.61\pm0.1$\\
\hline 
\end{tabular} 
\label{Table_optdep}
\end{table}

We compare our estimation to QSO's data with data from \citet{dallaglio08}, \citet{faucher08} , \citet{Becker13}, \citet{Becker15} and galaxy based optical depth estimation from \citet{Monzon20}. Our measurement from all our galaxies are in excellent agreement with each of these sample over the whole redshift range. At the high redshift end of our sample, the scatter of \citet{Becker15} also reaches our less/more transmitted galaxies. We fit the evolution of the optical depth for each of our sample with an analytical function of the form:

\begin{equation}
\label{func_teff}
    \tau_{eff} (z) = A\times (1+z)^{\gamma}
\end{equation}

The fit is carried out using the same algorithm as Sect.~\ref{results} using the scipy pacakge. We find, for the three sample:

\begin{itemize}
\item All galaxies: $A=2.655\times10^{-3}\pm0.001$, $\gamma=3.536\pm0.170$
\item More transmitted: $A=9.540\times10^{-6}\pm6.397\times10^{-6}$, $\gamma=6.717\pm0.413$
\item less transmitted: $A=0.026\pm0.006$, $\gamma=2.354\pm0.147$
\end{itemize}
 It is worth mentionning that the result of this fit is valid for the redshift range we consider here from $z_{source}$=2.85 to $z_{source}$=5.04.\\

\begin{figure}[h!]
\centering
\includegraphics[width=9.cm]{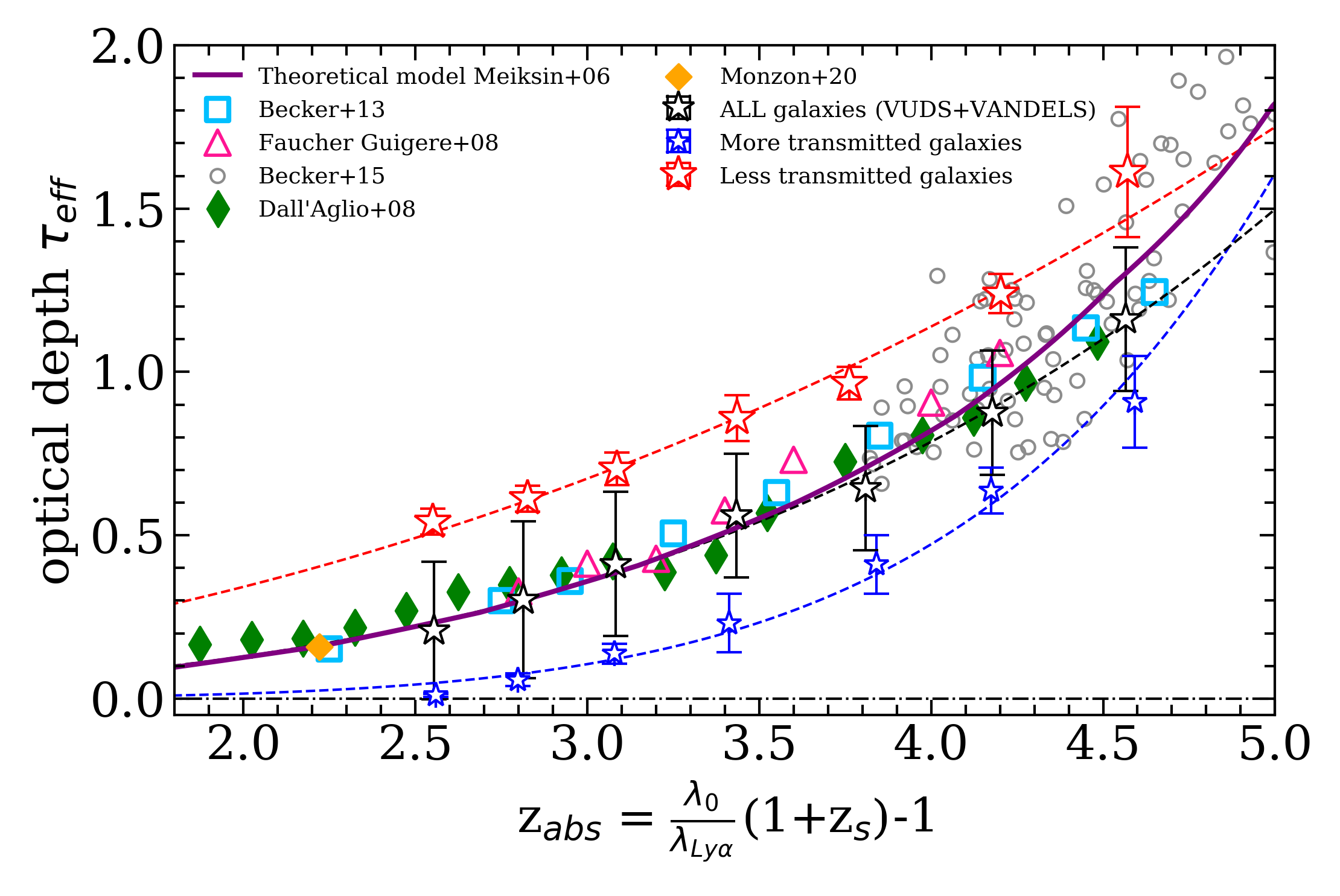}
\caption{Evolution of the optical depth for each sample of our galaxies and comparison with QSO's data. More IGM transmitted galaxies are shown by the blue stars, less transmitted galaxies are in red stars and black star represent the full sample of our galaxies. Dashed lines represent the fit to each sample and are in the same colors. We also show point from the literature with measurements from \citealt{dallaglio08} in green diamonds, from \citealt{faucher08} in pink triangles, from \citealt{Becker13} in light blue square and from \citealt{Becker15} in empty gray circles. The orange point is based on galaxy data from \citet{Monzon20}. Finally, the purple plain line shows the Meiksin modelisation (shifted to the $z_{abs}$, see Eq.~\ref{redtrans}).}
\label{opt_dep}
\end{figure}


Finally, as we can see, the evolution of the optical depth for both more/less transmitted galaxies is well separated at any redshift which would indicate that some galaxies may become visible, below the Ly$_{\alpha}$ line, at very different redshift depending only on their IGM optical depth. At a fixed optical depth, the difference in redshift reaches $\gtrsim 1$. 

\section{Discussion}
\label{disc}

\subsection{Asymmetry in the distribution of more/less transmitting lines of sight.}
\label{sec_asym}
As reported in section \ref{results}, we observe that in the highest redshift bins, the number of galaxies with a LOS classified as less transmitting is smaller than the the number of classified galaxies with a LOS classified as more transmitting. We investigate this asymmetry by looking  at the two different surveys independently (see Table \ref{asym_table}).

\begin{table}[h!]
\centering
\begin{tabular}{ccc}
\hline 
VANDELS & Ngal$_{more}$ & Ngal$_{less}$\\ 
\hline 
4.0$<$z$<$4.5 & 39 & 43  \\ 

4.5$<$z$<$4.8 & 18 & 14 \\ 

z$>$4.8 & 12 & 14  \\ 
\hline 
 VUDS & Ngal$_{more}$ & Ngal$_{less}$\\ 
\hline 
4.0$<$z$<$4.5  & 77 & 10 \\ 

4.5$<$z$<$4.8 & 40 & 2\\ 

z$>$4.8  & 33 & 4 \\ 
\hline 
\end{tabular} 
\caption{Number of galaxies with IGM transmission lower and higher than the theoretical mean from M06.}
\label{asym_table}
\end{table}

Taking into account the VANDELS survey alone the distribution of galaxies between less and more transmitting line of sights is not showing this asymmetry. On the other hand, when looking at VUDS data alone, the effect is present. This most probably indicates that the two pass fitting method is not completely efficient  when applied to the VUDS data. This couple be explained by the the number of points used for the dust estimation using photometric SED-fitting (we use up to 20 bands for VANDELS and around 10 for VUDS). This can be partially resolved by setting lower limits on the photometric fitting, i.e. forcing a lower dust attenuation as shown in T17. This would result in a more symmetrical distribution as for the VANDELS sources. The effect of such action would be to remove the slight jump that can be seen in Fig.~\ref{ev} at $z>4$ and shift the measurements of the optical depth closer to the Meiksin model (see Fig.~\ref{opt_dep}).


\subsection{$\gamma$ values}
As shown in the previous section, the fits of Fig.~\ref{opt_dep} show large differences in parameters range, especially for the $\gamma$ parameters that ranges from 2.3546 for the galaxies that are less transmitted to 6.3974 for the more transmitting galaxies. In the Lyman alpha forest studies framework, the $\gamma$ parameter is related to the evolution of the number density of Ly$\alpha$ forest clouds with redshift \citep{Press93}. In the same paper, the evolution of the optical depth is fitted by a functional of the form $\tau_{eff} (z) = A(1+z)^{\beta}$, with $\beta=\gamma+1$. In this case the $\beta$ value can be used directly to estimate the evolution of the number of clouds with redshift given by:

\begin{equation}
    \frac{dN}{dz} \propto B (1+z)^{\beta-1}.
\end{equation}
In that framework, we must remove 1 to all our values of $\gamma$ presented in Sect.~\ref{optdept}. We see that for the less transmitted galaxies, the $\gamma$ value is the lowest. This indicates a slower evolution with redshift. These line of sight are the one that are the most populated by clouds which leads to a lower transmission. This slower evolution can be explained by the fact that the strong increase of the optical depth at high-z \citep{Becker15} is less prominent for these line of sights because they are already heavily populated by clouds. 
On the contrary, the more transmitted galaxies are fitted with a high $\gamma$ value which indicates a strong evolution with redshift. For these galaxies, the lines of sight are less populated by clouds, leading to a higher IGM transmission. The strong evolution of the optical depth at high-z translates in a rapid increase of the number of clouds along the line of sight with redshift for the more transmitted galaxies.

\section{Conclusions}
\label{conc}

Using data from both VUDS and VANDELS programs, we presented in this letter the effect of the IGM at redshift z>2.7. Using averaged spectra we could observe the effect of the IGM for lines of sights that are more and less transmitting that the theoretical average prescription. This leads to two  important results:

\begin{itemize}
\item Firstly, the variance of the IGM is clearly visible from galaxy data. The flux for the sub-sample with the more transmitting lines of sight is on average 63\% higher in the 1070-1170\AA~region than the sub-sample with smaller IGM transmission. It has been already reported from the quasar point of view, e.g. \cite{Songaila04}. From the theoretical point of view, the large variance of the IGM was already shown by \citet{madau95} but not emphasized by more recent works. We defer a comprehensive comparison between our galaxy data with simulations to a future study. 
\item The estimation of the optical depth from the IGM transmission allows us to compare our data to multiple QSO studies and shows an excellent agreement.  Finally, the observed variance of the IGM described above could lead to different full-opacity redshift for different categories of lines of sight. The evolution of the optical depth from both more and less attenuated galaxies suggests that galaxies could start to be visible below Lyman alpha at very different redshift; at $z_s\sim6.16$ for less transmitted galaxies and at $z_s\sim6.80$ for the more transmitted galaxies. Nevertheless, QSOs studies at z>5 indicate that the evolution of the optical depth would become steeper as we increase redshift.

\end{itemize}


\begin{acknowledgements}
We wish to thank the anonymous referee for a careful reading of our manuscript and a insightful comments and corrections. We thank the ESO staff for their continuous support for the VANDELS survey, particularly the Paranal staff, who helped us to conduct the observations, and the ESO user support group in Garching. VUDS Data products are made available at the CESAM data center, Laboratoire d’Astrophysique de Marseille, France. RA acknowledges support from FONDECYT Regular Grant 1202007. All the plots have been made using the Photon software \citep{photon} and dfitspy \citep{dfitspy} was used for fits file handling.  We would like to dedicate this paper to the memory of Olivier Le F\`evre, PI of the VUDS survey and co-I of the VANDELS survey.
\end{acknowledgements}

\bibliographystyle{aa}
\bibliography{main.bib}


\end{document}